\documentclass[a4paper,11pt]{article}
\usepackage{pos}
\usepackage{outlines}
\usepackage{graphicx}
\usepackage{hyperref}
\usepackage{lineno}
\usepackage{placeins}
\usepackage{enumitem}

\let\OLDthebibliography\thebibliography
\renewcommand\thebibliography[1]{
  \OLDthebibliography{#1}
  \setlength{\parskip}{0pt}
  \setlength{\itemsep}{0pt plus 0.3ex}
}

\title{An update on site search activities for SWGO}
 \ShortTitle{SWGO Site Search Update}

\author{
\small{M. Santander$^{a,*}$, 
U.~Barres de Almeida$^{b}$, 
J.~A.~Bellido$^{c}$, 
T.~Bulik$^{d}$, 
C.~Dib$^{e}$, 
B.~Dingus$^{f}$, 
S.~Garcia$^{g}$, 
F.~Guarino$^{h}$, 
P.~Huentemeyer$^{i}$, 
D.~Mandat$^{j}$, 
E.~Meza$^{k}$, 
L.~Mendes$^{l}$, 
L.~Nellen$^{m}$, 
C.~Ocampo$^{n}$, 
L.~Otiniano$^{k}$, 
E.~Quispe$^{o}$, 
A.~Reisenegger$^{p}$, 
A.~C.~Rovero$^{q}$, 
F.~Sanchez$^{r}$, 
A.~Sandoval$^{s}$, 
R.~Yanyachi$^{t}$, 
H.~Zhou$^{u}$ 
on behalf of the SWGO Collaboration$^{w}$} \\
\normalfont{\small{
\emph{$^{a}$University of Alabama, Tuscaloosa, USA}, 
\emph{$^{b}$CBPF, Brazil}, 
\emph{$^{c}$University of Adelaide, Australia}, 
\emph{$^{d}$Astronomical Observatory, University of Warsaw, Poland}, 
\emph{$^{e}$CCTVal, Universidad Tecnica Federico Santa Maria, Chile}, 
\emph{$^{f}$University of Maryland, USA}, 
\emph{$^{g}$Universidad Nacional San Antonio Abad del Cusco, Peru}
\emph{$^{h}$Dipartimento di Fisica "E. Pancini" dell'Univerità degli Studi di Napoli and INFN Napoli, Italy}, 
\emph{$^{i}$Michigan Technological University, USA}, 
\emph{$^{j}$FZU, Czech Republic}, 
\emph{$^{k}$Comisión Nacional de Investigación y Desarrollo Aeroespacial, Peru}, 
\emph{$^{l}$LIP, Portugal}, 
\emph{$^{m}$Instituto de Fisica, UNAM, Mexico}, 
\emph{$^{n}$Atacama Astronomical Park, Chile}, 
\emph{$^{o}$Universidad de Moquegua, Peru}, 
\emph{$^{p}$Universidad Metropolitana de Ciencias de la Educación (UMCE), Chile}, 
\emph{$^{q}$Instituto de Astronomía y Física del Espacio (IAFE (CONICET-UBA)), Argentina}, 
\emph{$^{r}$Instituto de Tecnologías en Detección y Astropartículas (CNEA, CONICET, UNSAM), Argentina}, 
\emph{$^{s}$Instituto de Ciencias Nucleares, UNAM, Mexico}, 
\emph{$^{t}$Universidad Nacional San Agustín de Arequipa, Peru}, 
\emph{$^{u}$Tsung-Dao Lee Institute, Shanghai Jiao Tong University, China}.
}} \\
\normalfont{\small{\emph{$^{b}$ The Southern Wide-field Gamma-ray Observatory}}} \\
}

\emailAdd{jmsantander@ua.edu}

\abstract{The Southern Wide-field Gamma-ray Observatory (SWGO) is a project by scientists and engineers from 14 countries and 78 institutions to design and build the first wide-field, ground-based gamma-ray observatory in the Southern Hemisphere, with high duty cycle and covering an energy range rom hundreds of GeV to the PeV scale. The observatory will cover the Southern sky and aims to map the Galaxy’s large-scale emission, as well as detecting transient and variable phenomena. The host sites under consideration are at a minimum altitude of 4400 m.a.s.l. and comprise two types: flat plateaus of at least 1 km$^{2}$ for the installation of an array of tank-based water Cherenkov detectors (WCD), or large natural lakes for the direct deployment of WCD units. Four South American countries proposed excellent sites to host the observatory meeting these requirements. Argentina proposed two locations in the Salta province, Bolivia presented one site in Chacaltaya, Chile two locations within the Atacama Astronomical Park, and Peru two ground-based locations in the Arequipa district as well as lakes in the Cuzco region. The SWGO collaboration is currently conducting a site characterization study, gathering all the necessary information for site shortlisting and final site selection by the end of 2023. The process has reached the shortlisting phase, in which primary and backup sites for each country have been identified. The primary sites were visited by a team of experts from the collaboration, to investigate and validate the proposed site characteristics. Here we present an update on these site selection activities.

\vspace{4mm}

$^*$ Speaker

\ConferenceLogo{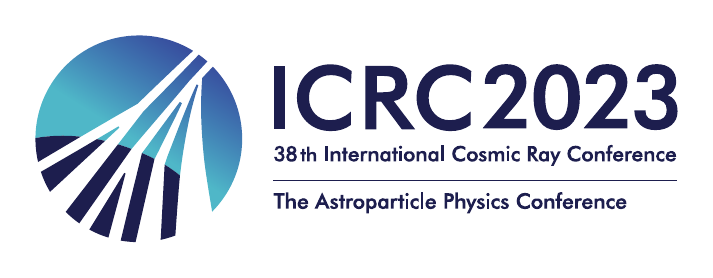}

\FullConference{%
38th International Cosmic Ray Conference (ICRC2023)\\
  26 July - 3 August, 2023\\
  Nagoya, Japan}
}

\begin{document}
\maketitle

\section{Introduction}\label{intro}

The Southern Wide-field Gamma-ray Observatory (SWGO)~\cite{Hinton:2021rvp} is a proposed observatory to be built in the Andes mountains of South America. SWGO will complement existing wide-field instruments currently operational in the northern hemisphere, such as HAWC and LHAASO, with the goal of observing gamma rays and cosmic rays from the southern sky with energies above $\sim 10$ GeV. The observatory will enable the study of the inner Galaxy, a direct observation of the Galactic Center, the observation of variable and transient sources such as active galactic nuclei, gamma-ray bursts and multimessenger triggers, as well as studies of fundamental physics.

\begin{figure}[h]
    \centering
    \includegraphics[width=\textwidth]{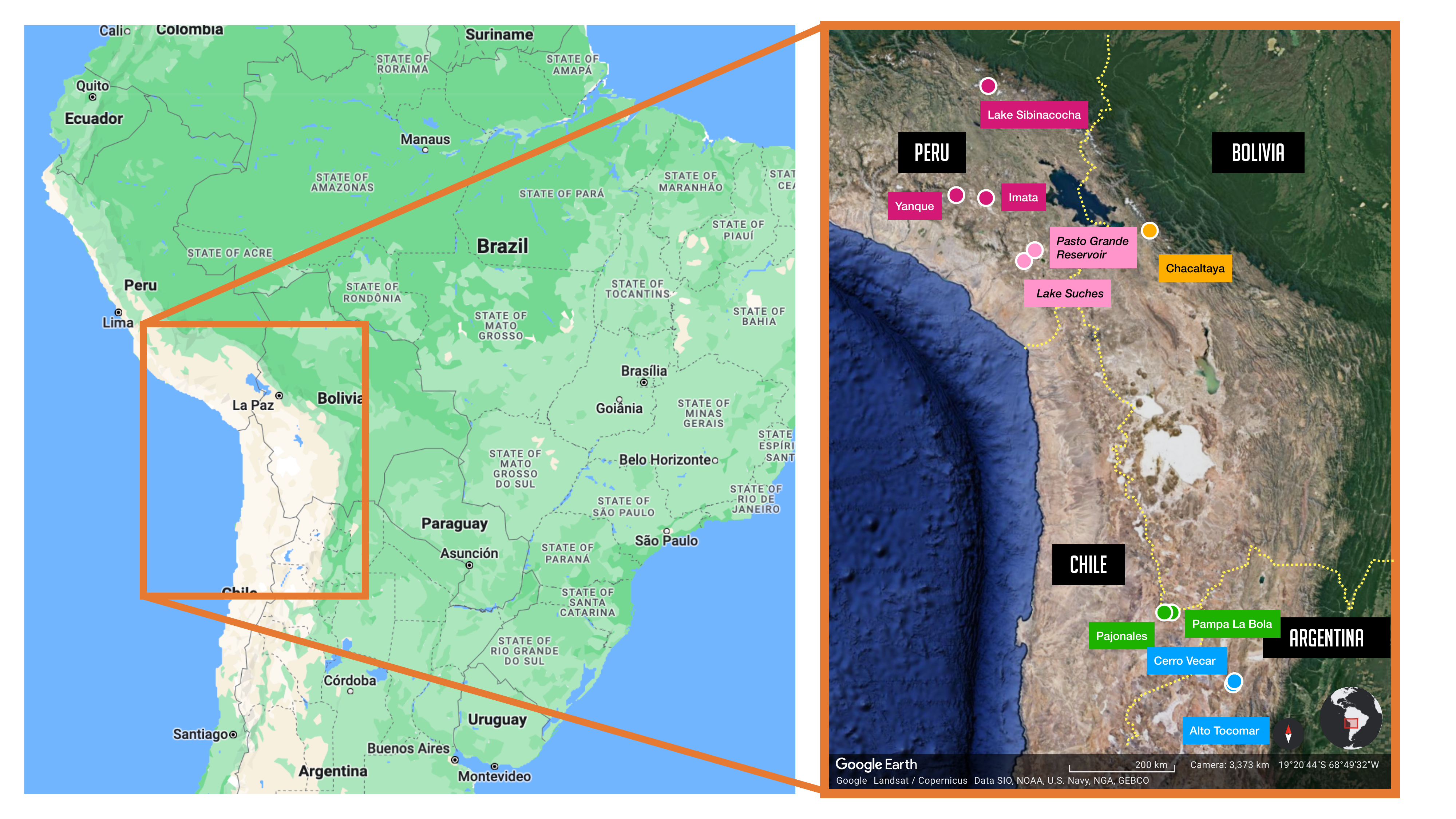}
    \caption{Map of sites currently under consideration for SWGO and their location in the continent.}
    \label{fig:south_am_map}
\end{figure}

The scientific and performance requirements of SWGO set constraints on the location and physical characteristics of the site that will host the detector. A set of minimum requirements were therefore defined by the SWGO collaboration to search for candidate sites:

\begin{itemize}[noitemsep]
\item \emph{Latitude}: The latitude of the site should be between 10$^{\circ}$ and 30$^{\circ}$ South in order to ensure a proper view of the Galactic Center region.
\item \emph{Altitude}: A minimum altitude of 4,400 m above sea level is required to provide a lower energy threshold, as air showers produced by gamma rays and cosmic rays in the desired energy range attenuate quickly in the atmosphere. 
\item \emph{Surface area}: The minimum area to host the SWGO reference configuration design requires 80,000 m$^{2}$ for an inner high fill-factor array, surrounded by an area of at least 200,000 m$^{2}$ for a sparser array. In order to accommodate potential future extensions of increase the high-energy reach of the instrument, an available area of $\sim$ 1 km$^{2}$ is required. 
\end{itemize}

These hard constraints (latitude, altitude, and available area) are the minimum requirements that should be met to identify a potential candidate site. The criteria for the selection of a site among those that pass the minimum requirements are the {\it performance}, {\it cost} and {\it risks} of constructing SWGO at a given site. The performance of the SWGO array at a given site could be affected by its physical characteristics, such as latitude (observable sky), altitude (energy threshold), and deviations from an ideal layout (trigger and acceptance considerations), among others. The costs involve both construction and operational aspects, while risks include environmental and social impacts (e.g. social, archaeological, or economical restrictions to the region; risk of earthquakes, flooding or heavy snow), economical and legal considerations (e.g. regulations regarding the importation of scientific equipment, site access, land use, etc.), and risks associated with on-site activities (e.g. access to required water, power, internet, safety conditions, health risks, local support, etc.). Given the scope of the project a site availability timescale of 20 years was selected as highly desirable.

There are currently three detector designs under consideration, all based on the water Cherenkov detection (WCD) principle proven by past and current observatories:

\begin{enumerate}[noitemsep]
\item \textbf{Tank design:} This is the baseline design. It consists of a large number of light-tight WCD tanks ($\sim$5,000) following a HAWC-like design \cite{hawc-web}. Each tank is about 3.8~m in diameter and 3~m in height and is equipped with one or more photon sensors overlooking a volume of purified water. The detector is composed of a core array with high fill factor surrounded by a sparser outer array. 
\item \textbf{Pond design:} This concept follows the design of previous and current instruments such as Milagro and LHAASO \cite{lhaaso-web}, where photon sensors are located in a large water pool (or "pond") instead of being isolated in individual WCD tanks. Photon sensors can still be optically isolated from each other by placing curtains or other types of barriers under water. This solution involves the construction of a large artificial pond at the detector site that will have to be completely covered to prevent light contamination.
\item \textbf{Lake design:} In this novel concept, a natural lake is used as part of the detector by placing the photo sensors into 'bags' that contain purified water and are immersed in the lake, keeping them at the right distance away from each other.
\end{enumerate}

The choice of detector technology and the site selection process are intimately linked, as some candidate sites may be able to host a particular detector design but not others. 

The site selection process started with the definition of representatives for the proposed host countries, a first round of data collection for each proposed site (including the technical and economical benefits of each), and the creation of a shortlist of candidate sites. The shortlist will include a {\it primary} and {\it backup} site for each detector technology that the country plans to host, where the preliminary primary and backup definitions are determined based on the relative merits of each site for a specific country. The information collected for the candidate sites is then collected into reports available to the entire collaboration. 

Four countries presented candidate sites to host SWGO, shown in Fig.~\ref{fig:south_am_map}\footnote{A map of all sites with additional photos and videos is shown \href{https://www.google.com/maps/d/edit?mid=1SXZPCuEFlhcMSpRys0uUEEu17z2LV0U&usp=sharing}{here}.}. We list them below along with their temporary designations as primary or backup sites: 

\begin{enumerate}[noitemsep]
\item \textbf{Argentina:} two sites were presented for a tank array: Cerro Vecar (primary) and Alto Tocomar (backup), both in the province of Salta, in the north of the country. 
\item \textbf{Bolivia:} One site was presented for a tank array: Cerro Estuquería near the existing ALPACA experimen~\cite{ALPACA}, in the Chacaltaya plateu. The site is currently under evaluation as a backup as it did not meet the minimum required area of 1 km$^{2}$. 
\item \textbf{Chile:} two sites were presented for a tank array: Pampa La Bola (primary) and Pajonales (backup), near the location of the ALMA observatory in the Atacama Desert.
\item \textbf{Peru}: two sites were presented for a tank or pond array: Yanque (primary) and Imata (backup) in the Arequipa region, and one site for a lake array: the lake Sibinacocha area in the Cuzco region (primary). Two additional lake sites are currently being evaluated: Pasto Grande and Suches, in the Arequipa region.  
\end{enumerate}

\begin{figure}
    \centering
    \includegraphics[width=0.8\textwidth]{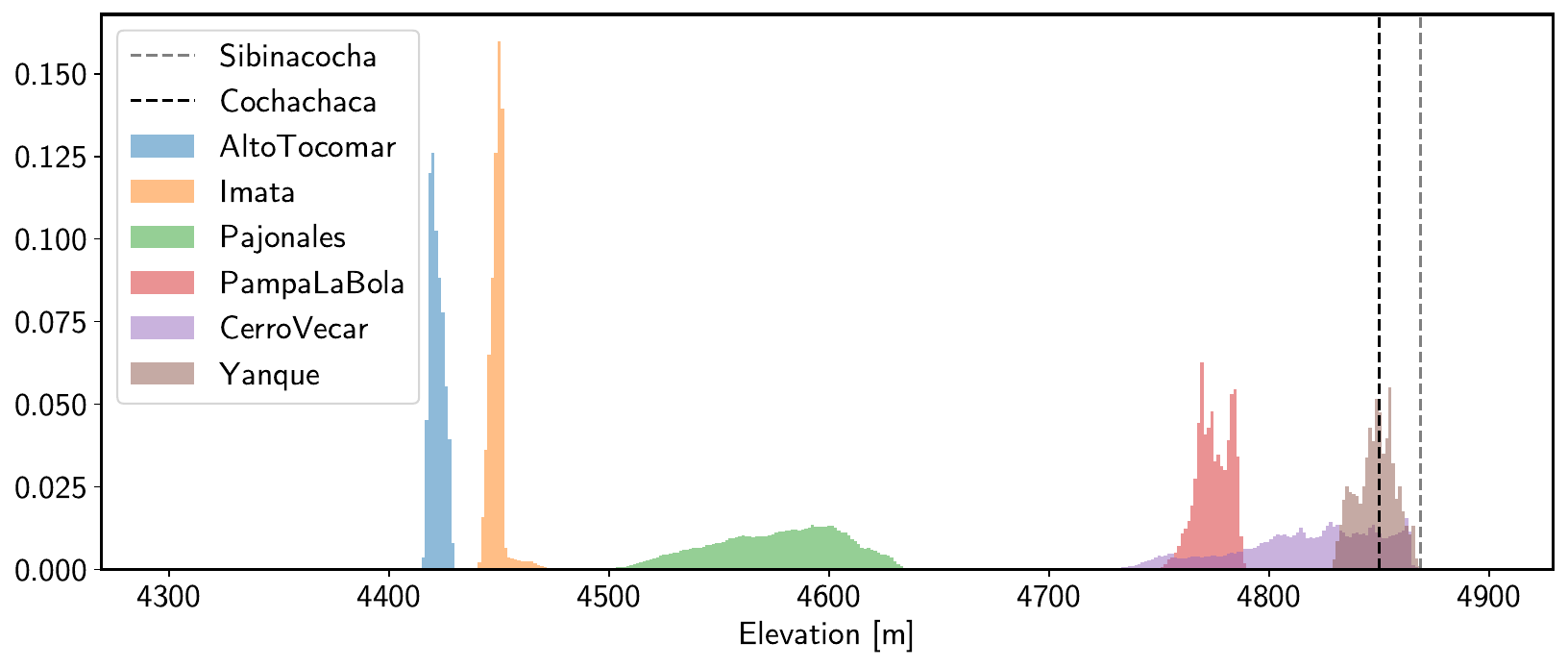}
    \caption{Histograms of distribution of altitudes of all sites visited. A narrow histogram means a flat site, and a wide histograms indicates a site with a large slope. All histograms have been adjusted to match the elevation from Google Earth.}
    \label{fig:hist-sites}
\end{figure}

The selection of a primary and backup site for each country and detector design during the shortlisting process is determined by the relative merits of each site within that country. Given the elevation difference between some sites (Fig.~\ref{fig:hist-sites}) within the same country, a simulation study is underway to help assess the impact on the scientific performance (in particular the low-energy sensitivity of the array) this could have to settle the primary or backup definition during the shortlisting step. 

A visit to the countries hosting at least one primary site (i.e. Argentina, Chile and Peru) was organized for October-November of 2022. In the next sections we describe the characteristics of the sites under consideration, including information collected during the 2022 visit. 

\section{Candidate sites in Argentina}\label{arg}

\begin{figure}[h!]
    \centering
    \includegraphics[width=\textwidth]{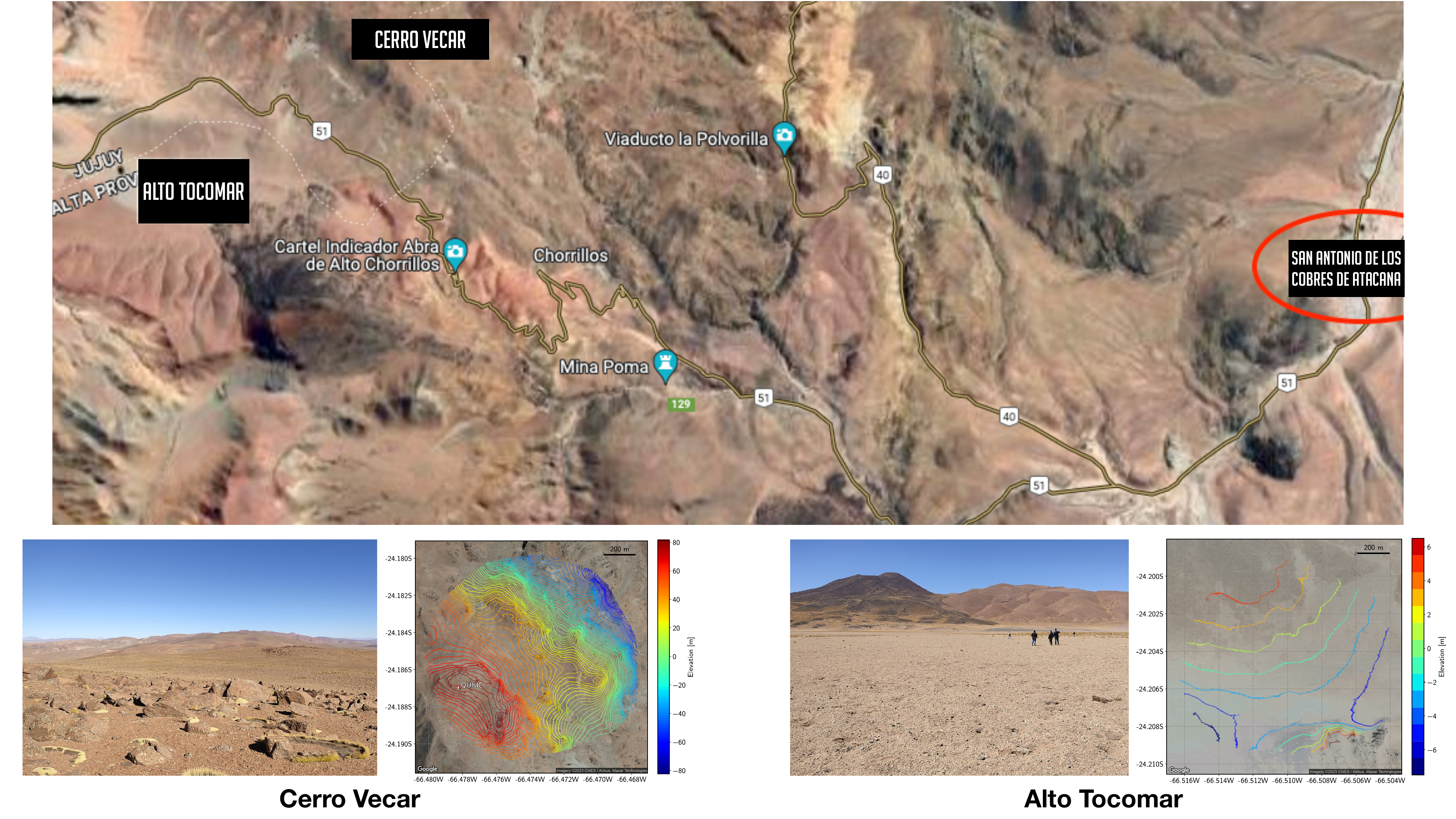}
    \caption{\emph{Top:} Map showing the location of both Argentinian sites with respect to SAC. \emph{Bottom left:} Photo of the Cerro Vecar site with a map showing elevation contours with 2 m spacing with respect to the average.  \emph{Bottom right:} Same as left for the Alto Tocomar site.}
    \label{fig:arg}
\end{figure}

The sites in Argentina (Fig.~\ref{fig:arg}) are 5 km away from each other, measured linearly, along National Road 51 (RN51) in the province of Salta. The nearest town is San Antonio de Los Cobres (SAC, pop. 5,480, 30 km away) accessible via an unpaved road. Both sites are proposed to host tank arrays. No electrical power and no internet connectivity exist at either site at the moment. The nearest illuminated internet fibre point is in San Antonio de Los Cobres. The average temperature on site is -5 °C (range: -15 °C to +10 °C) in the winter and +3 °C in the summer (-5 °C to +15 °C). 

The Cerro Vecar site ($\sim$4,800 m.a.s.l., 24.2° S) is near two observatories: QUBIC and LLAMA. The soil is sandy with some large rocks in the higher part of the area, and would have to be flattened prior to construction. The Alto Tocomar site (4,420 ma.s.l., 24.2° S) has sandy soil with no rocks and it has a flat area ($< 1$\% slope) that extends over 1 km$^{2}$. 

\section{Candidate site in Bolivia}\label{bol}

\begin{figure}[h!]
    \centering
    \includegraphics[width=0.7\textwidth]{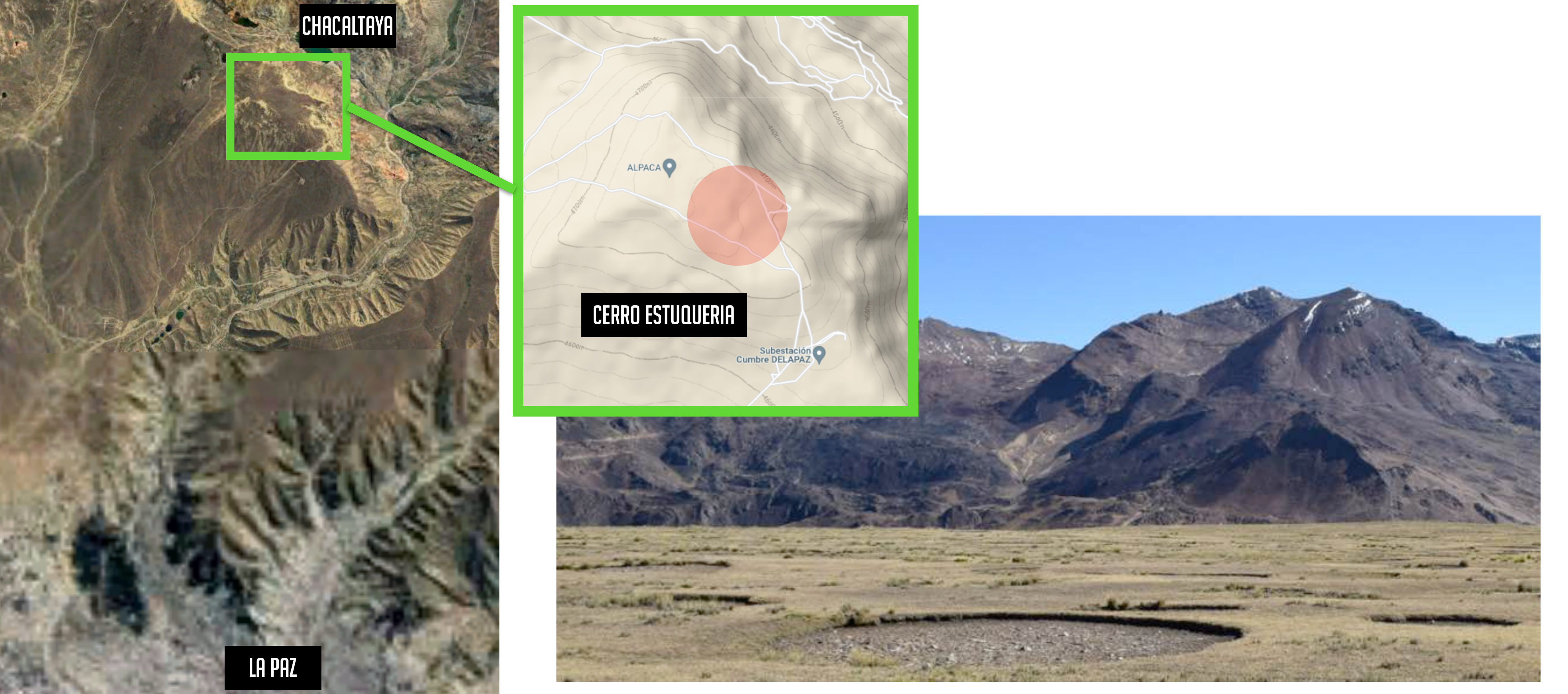}
    \caption{\emph{Left}: Location of the Cerro Estuqueria site with respect to La Paz, with the inset map showing a 600 m diameter circle representing a potential site location near ALPACA. \emph{Right:} View of the Chacaltaya plateau from ALPACA. }
    \label{fig:bol}
\end{figure}

A site in Cerro Estuqueria, Chacaltaya Plateau (4,750 m.a.s.l., 16° S, Fig.~\ref{fig:bol}), near the existing ALPACA observatory~\cite{ALPACA}, has been proposed to host a tank array. Due to the available area being 0.72 km$^{2}$, less than the minimum requirement, it is being considered only as a backup at the moment. The nearest cities are El Alto and La Paz (Bolivia’s capital), 15 km and 20 km away respectively measured linearly, connected to the site via an unpaved road. The average temperature in the Chacaltaya plateau is in the range between -4.2 °C and 5.5 °C. The site has no internet access at present, with the nearest internet access point about 10 km away. A 6.9 kV line runs near the site that could be used for electrical power. Water is not available on site, although underground water could be used as a source. 

% not mentioned water

\section{Candidate sites in Chile}\label{chi}

\begin{figure}[h!]
    \centering
    \includegraphics[width=\textwidth]{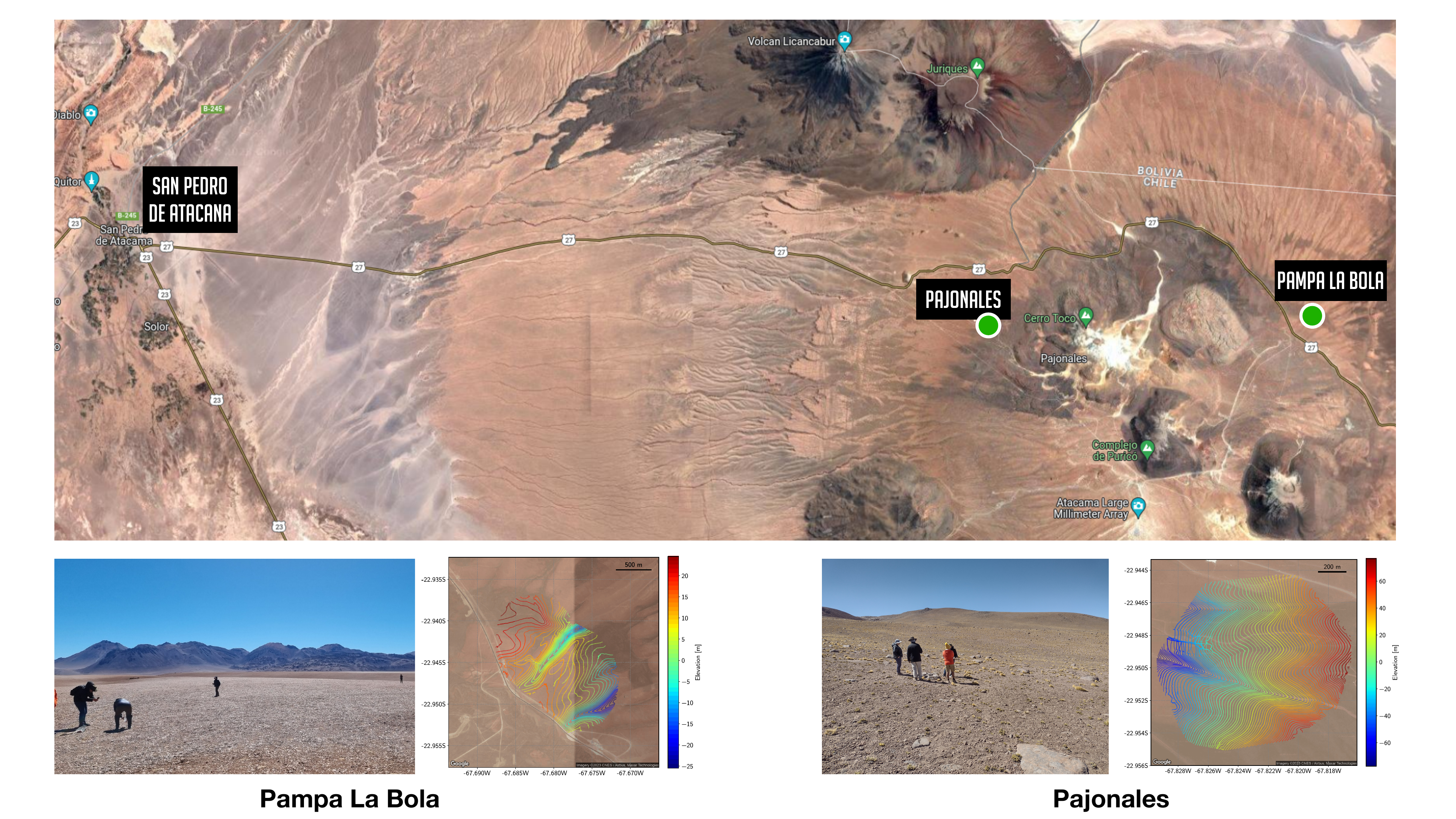}
    \caption{Map of sites currently under consideration for SWGO and their location in the continent.}
    \label{fig:chi}
\end{figure}

The Pampa La Bola (current primary, elevation 4,750 m.a.s.l., 22.9° S) and Pajonales (current backup, elevation 4,625 m.a.s.l., 23.0° S) sites are located within the Atacama Astronomical Park (AAP), about 18 km from each other along Road 27, a paved road that connects both sites with San Pedro de Atacama (pop. 4,000), about 45 km away. Both sites are proposed to host tank arrays. The average low temperature in the winter is about -18 °C, and the average high in the summer is +10 °C. 

No electrical power is available on site. Water is not available on site but it can also be purchased and transported from the nearby town of Calama, 150 km away. The nearest internet connection is the AAP fibre connection point at ALMA, 6 km away from Pampa La Bola. The soil in Pampa La Bola is made of very small stones and sand, with a flat area (slope $<$ 2\%) of several km$^{2}$, while in Pajonales there is a mixture of sand and a few rocks and the area has an overall slope of $\sim7$\%. The sites are near astronomical observatories at Cerro Toco and ALMA in the Chajnantor plateau (see Fig.~\ref{fig:chi}) .

\section{Candidate sites in Peru}\label{sites}

Two sites are proposed in Peru to host tank or pond arrays: Yanque (4,850 m.a.s.l., 15.7° S) and Imata (4,450 m.a.s.l., 15.9° S), both in the Arequipa region. Lows in the winter reach -16 °C, and highs in the summer are around 17 °C. Both regions are flat (slopes of $<3$\% for Yanque and $<2$\% for Imata), with an available area of 1 km$^{2}$ for Yanque, and multiple km$^{2}$ for Imata. No high-speed internet connection is available at either site. 

\begin{figure}[h!]
    \centering
    \includegraphics[width=\textwidth]{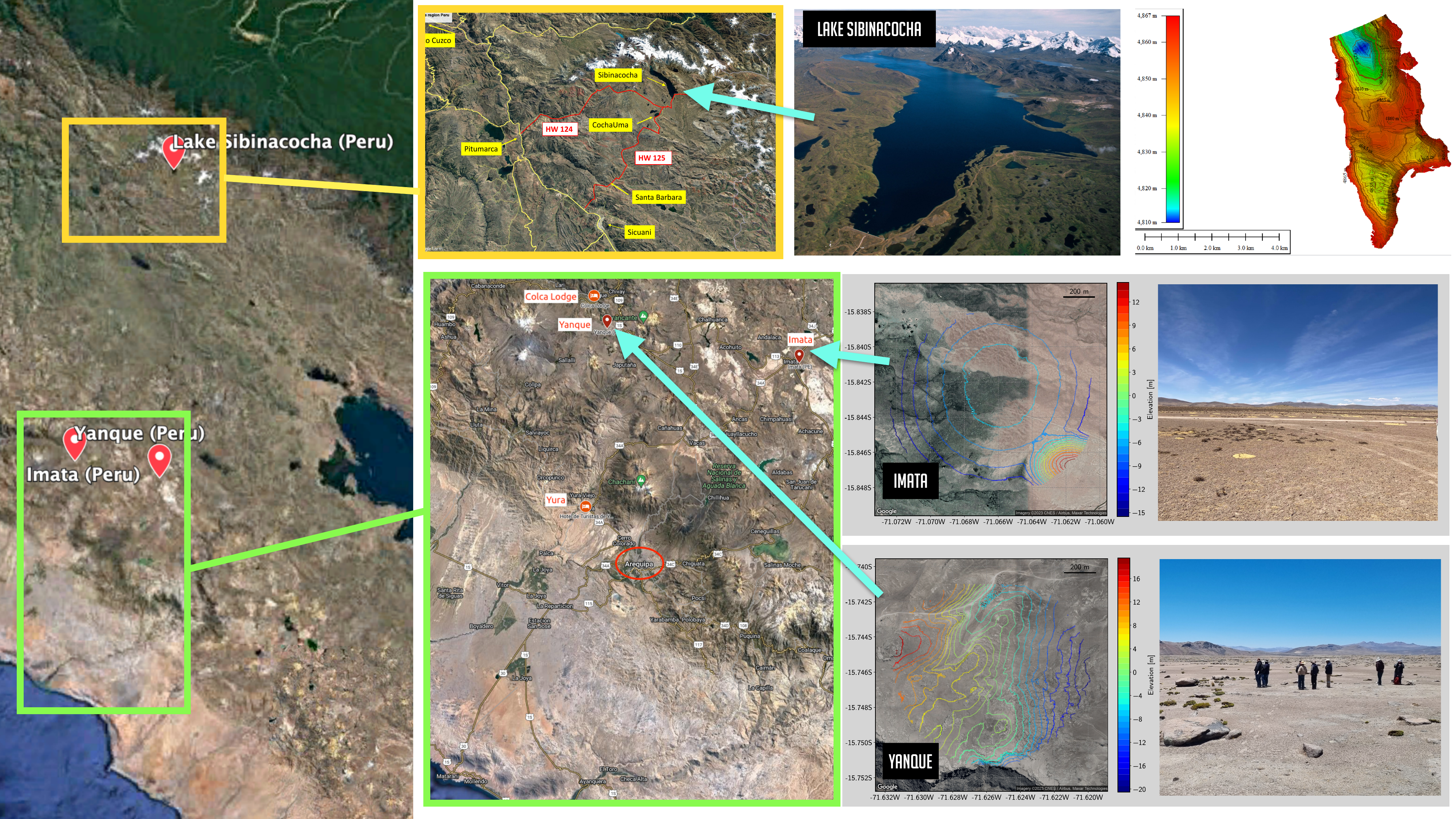}
    \caption{\emph{Left}: Map of the locations of different sites in Peru. \emph{Center:} regional maps for the sites in the Cuzco (top) and Arequipa (bottom) regions. \emph{Right:} Elevation contours (2 m spacing) for Imata and Yanque including photos of the sites taken during the 2022 visit (bottom) and a drone picture of lake Sibinacocha with bathymetry measurements of the southern part of the lake.}
    \label{fig:peru}
\end{figure}

For Yanque, the nearest town is Chivay (pop. 6,600), located about 30 km away (25 km on a paved road, and the last 5 km on a dirt road). Small lakes, used by the local community, are located nearby that could be used as a water source. Electrical power is available about 10 km away from the site. The soil is sandy with some big rocks are present. The Imata site is located 4 km away from the town of Imata (pop. 460), where electrical power is available. A river passing next to the site could be used as the water source, and the soil is sandy with very few rocks. 

Lake Sibinacocha (4,860 m.a.s.l., 13.9° S) is a very large lake with an area of 14 km$^{2}$ located in the Cuzco region. Two smaller lakes, Cochauma and Cochachaca, are located nearby. The area is accessible via unpaved roads and the nearest town is the village of Phinaya (pop. 340) about 2 km away from the nearest lakeshore. Bathymetry studies in the southern-most part of Sibinacocha yield an average depth of $\sim20$ m, and temperatures in the area are stable through the year, with highs of 10-14 °C and lows of -3 to -9 °C.  For more lake options, two additional sites in the Arequipa region (Pasto Grande and Suches) are currently under evaluation for a lake design detector.

\section{Engagement with local communities}

\begin{figure}
    \centering
    \includegraphics[width=\textwidth]{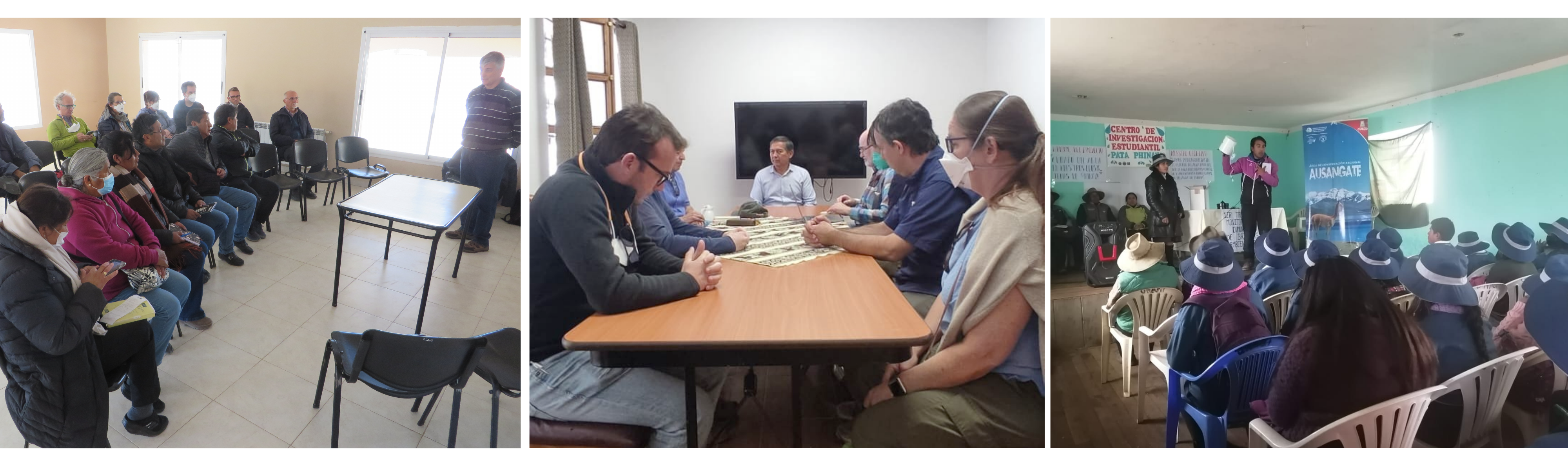}
    \caption{\emph{Left}: Presentation of the SWGO project to members of local pueblos originarios communities in San Antonio de los Cobres, Argentina, during the 2022 site visit. \emph{Center}: Meeting of the site visit team in 2022 with the mayor of San Pedro de Atacama, Chile, who is also the local representative of a pueblos originarios community. \emph{Right}: SWGO collaborator Dr. Sandro Arias has started a Science Center at the Phinaya school, Peru.}
    \label{fig:comm}
\end{figure}

One of the main goals of the site search process has been engaging local communities, and the native \emph{pueblos originarios} of the area, organizing meetings and outreach activities during our visits to the candidate sites.  
These meetings (Fig.\ref{fig:comm}) were very informative and have helped us understand their needs, requests, interests, and engagement better. 
More information on how to best engage and collaborate with local communities and pueblos originarios at each of the proposed sites is being collected, so that we can follow best practices of community participatory research and form sustainable and lasting partnerships that will help us better interact and integrate our activities with the community.

\section{Next steps}

The end of the shortlisting step is expected for the end of 2023, taking as input results from a study underway to evaluate the elevation-dependence of the scientific performance of the array. This will also help settle sites as either primary or backup options. The primary sites will then undergo full cost and environmental impact studies as well as risk assessments. Using this information, a preferred and backup site for the construction of SWGO will be selected in 2024.

% Bibtex references:
%\bibliographystyle{ICRC}
%\bibliography{main}

\providecommand{\href}[2]{#2}\begingroup\raggedright\endgroup

%% Full authors list (ONLY FOR COLLABORATIONS)

%\input{authorlist_SWGO.tex}
%
%\noindent \textbf{Note comment afterwards:} Collaborations have the possibility to provide an authors list in xml format which will be used while generating the DOI entries making the full authors list searchable in databases like Inspire HEP. For instructions please go to icrc2021.desy.de/proceedings or contact us under icrc2021proc@desy.de.\\
%
%\scriptsize
%\noindent
%first.author$^1$, 
%second.author$^2$, 
%third.author$^3$ % .... more names
%and 
%last.author$^{n}$ \\
%
%\noindent
%$^1$first.affiliation.
%$^2$second.affiliation. % .... more affiliation
%$^{m}$last.affiliation.

\end{document}